\documentclass[%
 reprint,
 amsmath,amssymb,
 aps,
]{revtex4-1}

\usepackage{graphicx}
\usepackage{dcolumn}
\usepackage{bm}
\usepackage{amsmath}
\usepackage{color} 
\usepackage{soul}

\begin{document}
\raggedbottom

\title{Sub-Doppler Cooling and Compressed Trapping of YO Molecules at $\mu$K Temperatures}%

\author{Shiqian Ding}
\email{shiqian.ding@colorado.edu }

\author{Yewei Wu}

\author{Ian A. Finneran}

\author{Justin J. Burau}

\author{Jun Ye}
\email{ye@jila.colorado.edu }

\affiliation{
JILA, National Institute of Standards and Technology and the University of Colorado, Boulder, Colorado 80309-0440 \\ 
Department of Physics, University of Colorado, Boulder, Colorado 80309-0390, USA
}

\date{\today}

\begin{abstract}
Complex molecular structure demands customized solutions to laser cooling by extending its general set of principles and practices. Yttrium monoxide (YO) has unique intramolecular interactions. The Fermi-contact interaction dominates over the spin-rotation coupling, resulting in two manifolds of closely spaced states, with one of them possessing a negligible Land\'e g-factor. This unique energy level structure favors dual-frequency DC magneto-optical trapping (MOT) and gray molasses cooling (GMC). We report exceptionally robust cooling of YO at 4 $\mu$K over a wide range of laser intensity, detunings (one and two-photon), and magnetic field. The magnetic insensitivity enables the spatial compression of the molecular cloud by alternating GMC and MOT under the continuous operation of the quadrupole magnetic field. A combination of these techniques produces a laser-cooled molecular sample with the highest phase space density in free space.
\end{abstract}

\maketitle

\section{Introduction}

Ultracold molecules~\cite{2009NJPReview, 2017ScienceReview} offer a new platform for quantum chemistry~\cite{Ospelkaus2010Science,TanyaPhotodissociation,Ni2019KRb}, strongly correlated quantum systems~\cite{Yan2013Nature}, quantum information processing~\cite{DeMille2002QL,Ni2018QL, Preskill2019,Oxford2019,Yiheng,CampbellHudson2018,CampbellHudson2019}, and precision tests of fundamental physics~\cite{2018EDM,Hutzler2017PRL,DeMillePV,ZelevinskyClock,2017JILAEDM,Flambaum2018PRL,HDIon,OdomTeH,SafronovaRMP}. There has been a long quest to create dense samples of ultracold molecules. Various molecular slowing methods have been proposed~\cite{BichrForce,ZeemanSlower,ZeemanSisyphus} and demonstrated~\cite{StarkDeceReview,MagneticDeceReview,RotatingDece,OpticalDece,RydbergDece,MicrowaveDece,CentrifugeDece}. Optoelectrical Sisyphus cooling has created polar molecular samples with submillikelvin temperatures~\cite{optoelectrical}. Ultracold diatomic molecules have been synthesized from ultracold atoms~\cite{KRb,RbCsNagerl,RbCsSimon,NaK,NaRb,NaCsLiu2018}, and very recently, a quantum degenerate gas of bialkali molecules has been realized~\cite{YeDegenerate}. Elastic atom-molecule interactions provide thermalization~\cite{Tobias2020} and collisional cooling \cite{KetterleCollision}.  Molecular ions have also been sympathetically cooled with co-trapped atomic ions, and quantum logic spectroscopy~\cite{QL2005,MolecularIonDing,MolecularIonLeibfried} has been successfully implemented ~\cite{MolecularIonWolf,MolecularIonChou,MolecularIonWillitsch}. 

Direct laser cooling of molecules provides a potentially species-independent approach for producing cold molecules. However, extending techniques used for laser cooling of atoms to molecules is challenged by the absence of cycling transitions associated with the complexity of molecular structure. To remove the large number of dark states in the form of rovibrational levels so that sufficient photon scatterings are ensured, cooling transitions should have maximally diagonal Franck-Condon factors~\cite{DiRosa2004} and optimized angular momentum selection rules~\cite{MOTproposal}. Following these initial proposals, the past few years have witnessed a rapid progress of laser cooling and trapping of molecules. Magneto-optical traps of diatomic molecules have been demonstrated for SrF~\cite{SrFMOT}, CaF~\cite{CaFMOTTarbutt,CaFDoyle}, and YO~\cite{2DMOT,our3DMOT}. 
Sisyphus-type cooling was reported using gray molasses cooling (GMC) for CaF and SrF to 50 $\mu$K ~\cite{CaFMOTTarbutt,moleculeODT,magtrapDemille}, and using magnetically assisted laser cooling for YbF to 100 $\mu$K~\cite{YbFcooling}. Triatomic molecules of SrOH and YbOH were cooled to $\sim$600 $\mu$K in one dimension~\cite{SrOH,YbOHDoyle}. GMC has been further enhanced with velocity-selective coherent population trapping, enabling deep cooling of CaF to $5~\mu$K~\cite{lambdaehanced,deepcooling}. Consequently, loading of molecules into optical dipole traps \cite{moleculeODT}, magnetic traps \cite{magtrapTarbutt,magtrapDemille}, and optical tweezer arrays \cite{moleculetweezer} has also been achieved. Other molecules that are currently under active investigations for direct laser cooling include BaH~\cite{BaHZelevinsky}, MgF~\cite{MgFYin}, BaF~\cite{BaFYan,BaFLangen}, AlF~\cite{AlFTruppe}, and CH~\cite{CHMcCarron}. 

In this work, we employ an effective cooling and trapping protocol that is specifically tailored for the unique structure of YO molecules. The presence of two independent electron-nuclear spin angular momentum states in the molecular ground state provides two quasi-independent paths for efficient GMC, along with robust dual-frequency DC magneto-optical trapping (MOT). The GMC-based approach provides the lowest laser-cooled molecular temperature of 4 $\mu$K over a broad range of laser tuning and magnetic field ($B$), such that exciting a Raman resonance does not bring further cooling, in contrast to previous reports~\cite{lambdaehanced,Lambda2013Li,Lambda2013KPRA,Lambda2013KEPL,Lambda2014Li,Lambda2016Na,Lambda2018Cs,Lambda2018Rb}. GMC is known to be magnetic field insensitive~\cite{Tarbutt2016NJP,BlueMOT,Tarbutt2018gray}, and this insensitivity is further enhanced in YO molecules by the existence of a ground state manifold with a negligible Land\'e g-factor. This feature enables the GMC to perform exceptionally well even at $B$ = 25 G, which is typically regarded as too large for sub-Doppler cooling to work~\cite{MagneticInsen,Tarbutt2016NJP}. Additionally, magnetically induced Sisyphus cooling~\cite{MagneticMetcalf,MagneticTannoudji,MagneticSrF,SrOH,YbFcooling} is observed under this field. This robust cooling mechanism against large $B$ has allowed us to demonstrate a novel scheme to significantly compress the molecular cloud by combining DC MOT trapping and sub-Doppler cooling.

\section{YO structure}

The excited state $A^2\Pi_{1/2}$ of YO follows a typical Hund's case (a) \cite{BrownBook}. The total angular momentum can be written as $\mathbf{F}=\mathbf{J}+\mathbf{I}$, where $\mathbf{J}$ is the total angular momentum excluding the nuclear spin $\mathbf{I}$.
The ground state $X^2\Sigma^+$ belongs to Hund's case (b), but with an extraordinarily large Fermi contact interaction (772 MHz), which makes it a special $b_{\beta S}$ case \cite{BrownBook}. The electron spin $\mathbf{S}=1/2$ and the yttrium nuclear spin $\mathbf{I}=1/2$ are coupled by Fermi contact interaction, forming an intermediate $\mathbf{G}=\mathbf{S}+\mathbf{I}$, which is further coupled with molecular rotation $\mathbf{N}$ by an electron spin-rotation interaction to form $\mathbf{F}=\mathbf{G}+\mathbf{N}$.

For $X^2\Sigma^+$, a considerable magnetic moment comes only from the electron spin. For the G=0 F=1 manifold, the  Land\'e g-factor is 0.01 in the weak field limit, where $B$ is not sufficiently strong (hundreds of G) to decouple the electron spin from the nuclear spin~\cite{MarkYeoThesis}. 

The cooling transition employs the $X^2\Sigma^+(N=1) \xrightarrow{} A^2\Pi_{1/2} (J'=1/2)$ transition~\cite{2DMOT,YOslowing,our3DMOT} at 614 nm. 
With highly diagonal Franck-Condon factors, only two vibrational repumpers at 648 nm and 649 nm are required to scatter enough photons out of $v$=1 and 2 states.
Compared with other laser-cooled molecules~\cite{SrFMOT, CaFDoyle,CaFMOTTarbutt}, YO has an extra decay channel associated with an intermediate $A'^2\Delta_{3/2}$ state, which decays to $X^2\Sigma^+(N=0,2)$.
We optically repump the $X^2\Sigma^+(N=2)$ state (through $A^2\Pi_{1/2} (J'=3/2)$) to $X^2\Sigma^+(N=0)$ and mix the latter with $X^2\Sigma^+(N=1)$ using microwaves to bring the molecules back to the cycling transition~\cite{our3DMOT}.  

\section{DC MOT}

\begin{figure}[tb]
\centering
\includegraphics[width=1.0 \columnwidth]{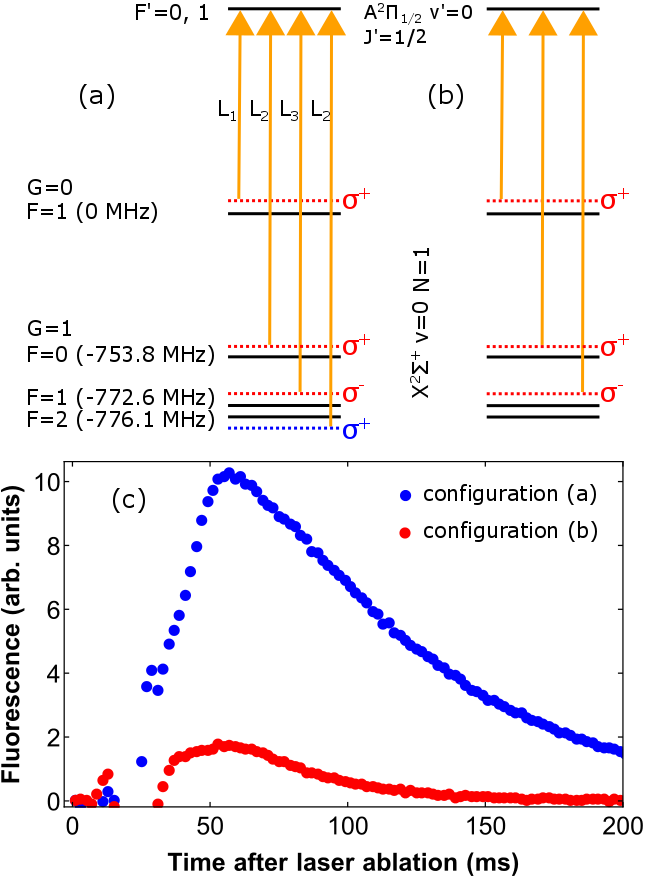}
\caption{\label{fig:DCMOT} 
DC MOT. (a) The optimal polarization settings for all the hyperfine components for the dual-frequency MOT. For illustration purpose, we show the polarization of the beam propagating along the $-z$ direction, and the magnetic field points along $+z$ at positions $z>0$ \cite{Tarbutt2015dualFreq,ImprovedSrFMOT}. The hyperfine structure of the excited manifold
is not resolved. (b) The optimal polarization settings for the red-detuned MOT. (c) Differential MOT fluorescence between two opposite quadrupole field directions, for the laser configurations shown in panels (a) and (b). We switch on a magnetic field gradient of 12 G/cm at 23 ms, and the loading process takes about 35 ms.
}
\end{figure}

\begin{figure*}[tb]
\centering
\includegraphics[width=2.0000 \columnwidth]{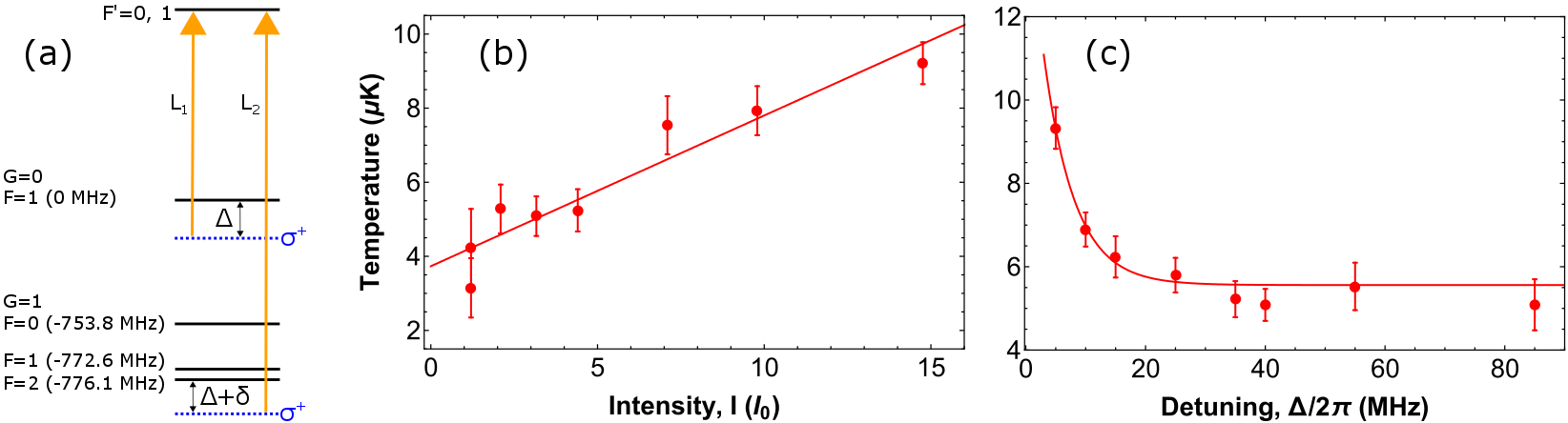}
\caption{\label{fig:TVsPara}
Characterization of GMC. (a) Configuration of laser detunings. (b) Temperature versus intensity $I$ with single-photon detuning $\Delta=8.3 \Gamma$ and Raman detuning $\delta=0$ between the G=0 F=1 and the G=1 F=2 manifolds. The solid line is a linear fit. (c) Temperature versus $\Delta$ with $I=3.2 I_0$ and $\delta=0$. 
The solid line is a guide for the eye. The error bars in the plots correspond to $1\sigma$ statistical uncertainty. 
}
\end{figure*}

Our systematic investigation of sub-Doppler cooling of YO begins with a DC MOT~\cite{Tarbutt2015NJP,Tarbutt2015dualFreq} loaded with the buffer-gas-cooled and laser-slowed YO molecules~\cite{YOslowing,our3DMOT}. Depending on the hyperfine states, shown in Fig.~\ref{fig:DCMOT}(a), the optimal laser frequency detunings and polarization settings vary accordingly to provide robust trapping forces arising from a dual-frequency MOT~\cite{Tarbutt2015dualFreq}, a red-detuned type-I (F$^{\prime}$ $>$ F) MOT, or a red-detuned type-II (F$^{\prime}$ $\leqslant$ F) MOT~\cite{Tarbutt2015NJP}. Here F$^{\prime}$ and F represent the angular momentum quantum number in the excited and ground states, respectively.

Considering its robust performance, we start our discussion with the dual-frequency MOT. To generate the laser beams, we split the main cooling laser into three beams. Beam one ($L_1$) is red detuned from G=0 F=1, and the other two beams ($L_2$ and $L_3$) pass through separate acousto-optic modulators (AOMs) to address the G=1 manifold. The AOM for $L_2$ is driven with dual RF frequencies to produce optical fields red detuned from G=1 F=0 and blue detuned from G=1 F=2. $L_2$ is overlapped with $L_1$ via a non-polarizing beamsplitter. $L_3$ has opposite polarization and is sent through the second AOM to address G=1 F=1. Finally, $L_1 / L_2$ and $L_3$ containing four frequency components of roughly equal power are combined via a polarizing beamsplitter, and further merged with the two vibrational repumpers using dichroic mirrors. After spatial mode filtering through a polarization maintaining optical fiber, the output is sent to the trapping region through three perpendicular axes and is retroreflected at the end. The MOT beam has a $1/e^2$ diameter of 11 mm, and is slightly converging to compensate for optical loss due to propagation through windows and wave-plates.

The dual-frequency MOT lifetime decreases with the increasing MOT beam intensity, which suggests a loss mechanism of optical pumping to dark states. We set a nominal intensity $I$ = 1.4$I_0$ for the DC MOT, where $I_0$=2.7 mW/cm$^2$ is the estimated saturation intensity. At this intensity, the detunings for the hyperfine components, from top to bottom in Fig.\ref{fig:DCMOT}(a), are tuned to be -5.8 MHz, -9 MHz, -5.8 MHz and +2.6 MHz, respectively, to maximize the number of trapped molecules, resulting in a MOT lifetime of 90 ms. The molecular sample has a $2\sigma$ diameter of 2.8 mm (3.9 mm) along the axial (radial) direction and a temperature of 2 mK. The number of trapped molecules is 1.1$\times$10$^5$, seven times that reported in our previous work~\cite{our3DMOT}. This increase comes from a number of optimizations involving the trapping and repumping lasers, the buffer-gas cell, and the vacuum. We find it important to also address the $X^2\Sigma^+(N=3)$ state. Otherwise, the number of trapped molecules is reduced by $\times$2, and the lifetime is shortened to 48 ms. The $X^2\Sigma^+(N=3)$ repumper is generated by sending the $X^2\Sigma^+(N=2)$ repumper through an electro-optic modulator driven at 9.79 GHz. 

The dual-frequency MOT has contributions from the other two trapping mechanisms, red-detuned type-I and red-detuned type-II MOT~\cite{Tarbutt2015NJP,Tarbutt2015dualFreq}. To determine their relative contributions, we disable the blue detuned component in $L_2$, shown in Fig.~\ref{fig:DCMOT}(b), and study the consequences on the MOT performance. As shown in Fig.~\ref{fig:DCMOT}(c), without the blue component the MOT produces five times fewer molecules and half of the lifetime as compared with the dual-frequency MOT. This clearly demonstrates that the dual-frequency trapping mechanism dominates the trapping force for the MOT~\cite{Tarbutt2015dualFreq}.

The YO dual-frequency MOT further benefits from the close spacing (3.5 MHz) of G=1 F=1 and G=1 F=2, in comparison to the cooling transition linewidth of $\Gamma \sim 2\pi \times 4.8 ~\text{MHz}$ \cite{YOLifetime}. Therefore, G=1 F=1 and G=1 F=2 both contribute to the dual-frequency trapping mechanism, increasing the overall trapping efficiency. This may be responsible for our observation that the DC MOT traps $\sim$20$\%$ more molecules than the RF MOT~\cite{our3DMOT}, contrary to the results reported for CaF~\cite{CaFDoyle} and SrF~\cite{SrFRF}.

We note that the red-detuned MOT (Fig.~\ref{fig:DCMOT}(b)) already employs the correct polarization for both laser components addressing the G=1 manifold. For the two closely spaced states of G=1 F=1,2, we find trapping for G=1 F=2 and anti-trapping for G=1 F=1~\cite{Tarbutt2015NJP}. Our result shows that the force from G=1 F=2 overwhelms that from G=1 F=1, which is consistent with the observation in Ref.~\cite{ImprovedSrFMOT} for SrF molecules.

\section{Gray Molasses Cooling}
Once the DC MOT is loaded, we blue detune $L_1$, remove the red-detuned component of $L_2$, and switch off $L_3$ and the quadrupole magnetic field for gray molasses cooling~\cite{GMCHansch,GMC,Tarbutt2016NJP}. We keep the vibrational repumpers on and switch off microwave field. GMC works on type-II transitions, and relies on position-dependent dark states that are formed from the spatial variation of the laser polarization and intensity. Molecules optically pumped into dark states can come out to a bright state via motion-induced non-adiabatic transitions~\cite{GMCHansch}. On average the spatially varying ac Stark shift of the bright state gives rise to motional damping more often than acceleration, leading to Sisyphus-type cooling~\cite{GMCHansch}. 

Unlike the hyperfine structure of SrF and CaF, YO has a large frequency gap between G=0 and G=1, shown in Fig.~\ref{fig:TVsPara}(a). As a result, the two laser components, $L_1$ and $L_2$ independently address these two manifolds, unless they are tuned on a Raman resonance ($\delta$ = 0). After 8 ms of GMC, we determine the temperature of the molecular cloud using standard ballistic expansion measurements. For a wide range of parameters that we have explored in this work, the molecular temperature remains robustly below 10 $\mu$K (Fig.~\ref{fig:TVsPara}(b) and (c)).  To make an accurate measurement of such low temperatures, we compress and cool the cloud first using a novel scheme, described in detail below. Unless otherwise specified, we plot the average temperature $T_{avr}=T_r^{2/3} \times T_a^{1/3}$, where $T_r$ and $T_a$ are the temperatures along the radial and axial directions, respectively. 

\begin{figure}[t]
\centering
\includegraphics[width=1.0 \columnwidth]{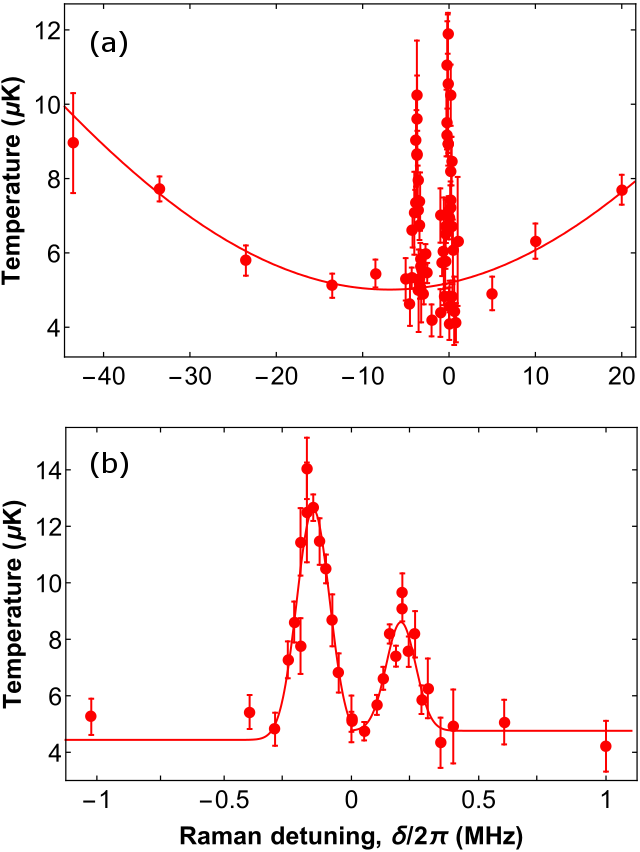}
\caption{
\label{fig:TVsRaman}
Scan of $\delta$ in GMC. (a) Temperature versus $\delta$, with $I=3.2 I_0$ and single-photon detuning $\Delta=8.3\Gamma$ for G=0 F=1. The temperature is elevated near two Raman resonance conditions $\delta=0$ (for G=1 F=2) and $\delta/2\pi=-$3.5 MHz (for G=1 F=1). The solid line is a guide for the eye.
(b) Zoomed  -in scan near the resonance condition $\delta=0$, with $I=3.2I_0$ and $\Delta=12.5\Gamma$. The solid lines are Gaussian fits.
}
\end{figure}

The YO temperature depends on the molasses beam intensity and frequency detuning in a way consistent with theory~\cite{Tarbutt2016NJP}. As shown in Fig.~\ref{fig:TVsPara}(b) and (c) for $\delta$ = 0, the temperature increases linearly with the intensity, and decreases with the detuning before settling down to a constant. The momentum diffusion coefficient $D_p$ increases (decreases) with intensity (detuning), while the damping coefficient $\alpha$ is both intensity and detuning insensitive~\cite{Tarbutt2016NJP}. The temperature is determined by $k_BT=D_p/\alpha$, where $k_B$ is the Boltzmann constant. It is worth mentioning that for intensity smaller than the minimum value shown in Fig.~\ref{fig:TVsPara}(b), the molasses becomes too weak to hold molecules against gravity.

This behavior is similar to the sub-Doppler cooling with type-I transitions~\cite{Tannoudji1989}, where $T\propto|\Omega|^2/\Delta$. Here $\Delta$ is the detuning and $\Omega$ is the Rabi frequency, which is proportional to $\sqrt{I}$.  

To understand the performance of GMC around a Raman resonance, we next investigate the temperature dependence on the Raman detuning $\delta$, while keeping $I$ = 3.2$I_0$ and the single-photon detuning $\Delta/2\pi$ for G=0 F=1 fixed at +40 MHz (8.3$\Gamma/2\pi$). The cooling is insensitive to $\delta$ over an extensive range, as shown in Fig.~\ref{fig:TVsRaman}(a). 
For $\delta/2\pi$ = -43.5 MHz, the remaining component of $L_2$ is on resonance with G=1 F=1. Thus, cooling (to 9 $\mu$K) arises purely from the G=0 F=1 manifold. Similarly, if we set $L_1$ on resonance with G=0 while having $L_2$ blue detuned from G=1, the cooling temperature reaches as low as 15 $\mu$K. We thus conclude that both the G=0 and the G=1 manifolds are responsible for GMC, and their combination leads to efficient sub-Doppler cooling of molecules to the lowest temperature while being robust against the Raman detuning~\cite{Bartolotta2020}.

When $\delta$ is tuned slightly away from 0, we observe striking rises of temperature, followed with quick decreases back down to 4 $\mu$K when $|\delta|$ becomes sufficiently large of a few hundred kHz. The two resonance-like features are symmetrically located around $\delta$ = 0 as shown in Fig.~\ref{fig:TVsRaman}(b). Ref.~\cite{lambdaehanced} reports a similar rise of temperature away from $\delta$ = 0 for CaF. For YO, beyond the two peaks the temperature returns quickly to the same value as that on the Raman resonance. The observed peak width and splitting between the two temperature peaks increase (decrease) with the laser intensity (the single-photon detuning $\Delta$). Our understanding for this effect is the following. GMC depends on transient dark states~\cite{Tarbutt2016NJP,Tarbutt2018gray}, and generally, a less stable dark state causes more momentum diffusion and leads to a higher temperature.  Near the Raman resonance condition, dark states formed in both G=0 F=1 and G=1 F=2 can be destabilized by their cross coupling. It leads to enhanced photon scattering and deteriorated cooling over a range with the width on the order of the two-photon Rabi frequency $\Omega_R$. This is different from the $\delta$ = 0 situation where a new stable dark state is formed with the superposition of Zeeman sublevels from different hyperfine states~\cite{lambdaehanced,deepcooling}, resulting in a $\Lambda$-type GMC.   

\begin{figure}[t]
\centering
\includegraphics[width=1.0 \columnwidth]{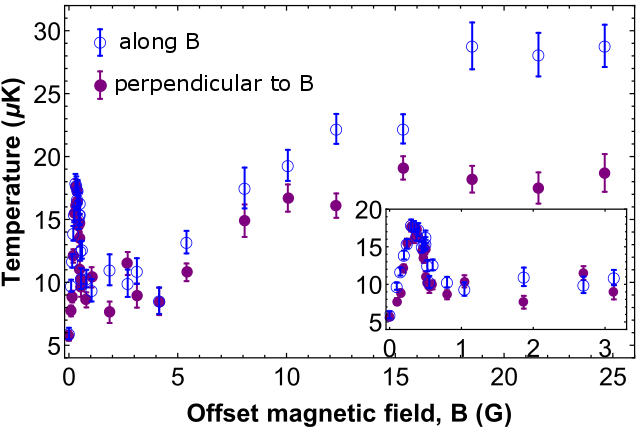}
\caption{\label{fig:TVB}
Temperature under a uniform $B$ applied along $z$, with $I=3.2 I_0$, $\Delta=8.3\Gamma$, and $\delta=0$.
The blue open circles (purple filled points) represent the temperature measured along (transverse to) $B$. The inset shows the temperature from 0 G to 3 G.}
\end{figure}

A similar temperature increase is observed for the applied microwave coupling between the N=0 and the N=1 states. We thus switch off the microwaves for state remixing during GMC, otherwise, the achieved temperature is always $>$15 $\mu$K. Note that the $\sim$15$\%$ of molecules initially populated in N=0 states are lost.

\begin{figure*}[t]
\centering
\includegraphics[width=2.0 \columnwidth]{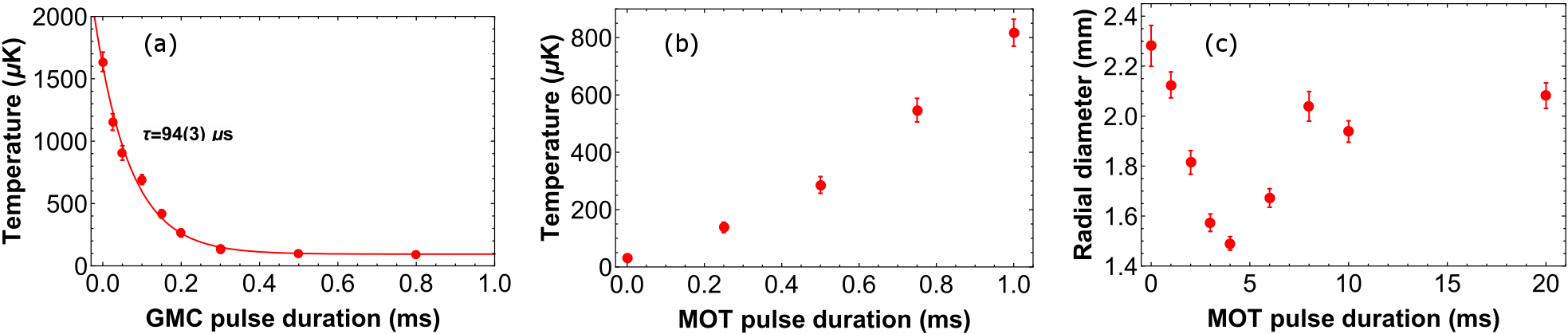}
\caption{\label{fig:compression}
Compression of the molecular cloud by combining the GMC with the dual-frequency DC MOT at a $B$ gradient of 47 G/cm. (a) Temperature versus the duration of the GMC pulse, with $I=30I_0$, $\Delta=3.1\Gamma$, and $\delta=0$. The solid line is an exponential fit. (b, c) Temperature and $in$-$situ$ diameter of the molecular cloud versus the duration of a single MOT pulse, which is applied after the 2 ms GMC pulse.
}
\end{figure*}

\section{GMC under $B$}

Considering the negligibly small Land\'e g-factor for G=0, it is interesting to explore how well GMC performs in the presence of $B$. We apply a magnetic field along the axial direction and measure the temperatures both along and perpendicular to $B$ with $\delta=0$. Figure~\ref{fig:TVB} demonstrates that the GMC is surprisingly robust over a large range of $B$ values, along with a few remarkable features.

First, the temperature appears to have a pronounced peak near 0.3 G. This resonance-like feature is attributed to the Zeeman shifts of the G=1 F=2 sublevels that destabilize the dark state formed both in its own manifold through Larmor precession, and in the G=0 F=1 manifold through Raman coupling. 
The temperature dependence on $B$ then becomes flat between 0.6 G to 4 G, regardless of whether $L_2$ is tuned on resonance with G=1 F=2 or remains on Raman resonance with $L_1$. This observation reveals that, within this range of $B$, GMC arises mostly from G=0 F=1, which has a small Land\'e g-factor and thus forms a relatively stable dark state.

At $\sim$4 G, the temperature starts to rise with $B$, indicating that the dark state in G=0 F=1 becomes unstable. The temperature measured in the direction perpendicular to $B$ rises slower than that along $B$, and reaches a plateau after $\sim$10 G. This effect is attributed to the magnetically induced Sisyphus cooling~\cite{MagneticMetcalf,MagneticTannoudji,MagneticSrF,SrOH,YbFcooling}, after the original GMC becomes less effective. The laser field pumps YO into a dark state, which is remixed with a bright state via Larmor precession. This process damps the molecular motion in a Sisyphus fashion~\cite{MagneticTannoudji}. In the direction transverse to $B$, the spatially varying dark/bright state mixing is much stronger than that along $B$, resulting in lower temperatures. 
The magnetically induced Sisyphus cooling works at $\sim$25 G for YO, substantially larger than any other atoms or molecules~\cite{MagneticMetcalf,MagneticTannoudji,MagneticSrF,SrOH,YbFcooling}, where the field applied is nominally $\sim$1 G. Again, such an exceptional magnetic insensitivity must be attributed to the small Land\'e g-factor of G=0 F=1. 

\section{Compression}

Such exceptionally robust GMC against the magnetic field brings an intriguing capability. In the presence of the quadrupole magnetic field used for the MOT, molecules can still be cooled via GMC to a temperature much lower than that for the MOT. Thus, with the quadrupole field turned on continuously, we can alternate the laser beams between MOT and GMC parameters so as to apply the strong trapping force of the MOT to the ultralow temperature molecular sample prepared by GMC.  This scheme works effectively as the GMC process is fast relative to MOT heating, such that the compression from the MOT dominates over the expansion during the GMC.

We measure the cooling (heating) speed of GMC (MOT) by recording the temperature for different GMC (MOT) pulse durations, as shown in Fig.~\ref{fig:compression}(a, b). The cooling speed (1/$\tau$) increases with the laser intensity and is not sensitive to the detuning between 1$\Gamma$ to 6.2$\Gamma$. With the full laser power (30$I_0$) and detuning 3.1$\Gamma$, molecules are cooled from 2 mK to 100 $\mu$K with a $1/e$ time constant $\tau\sim$ 94 $\mu$s. The MOT heating process, on the other hand, takes $\sim$1 ms. 

\begin{figure}[t]
\centering
\includegraphics[width=1.0 \columnwidth]{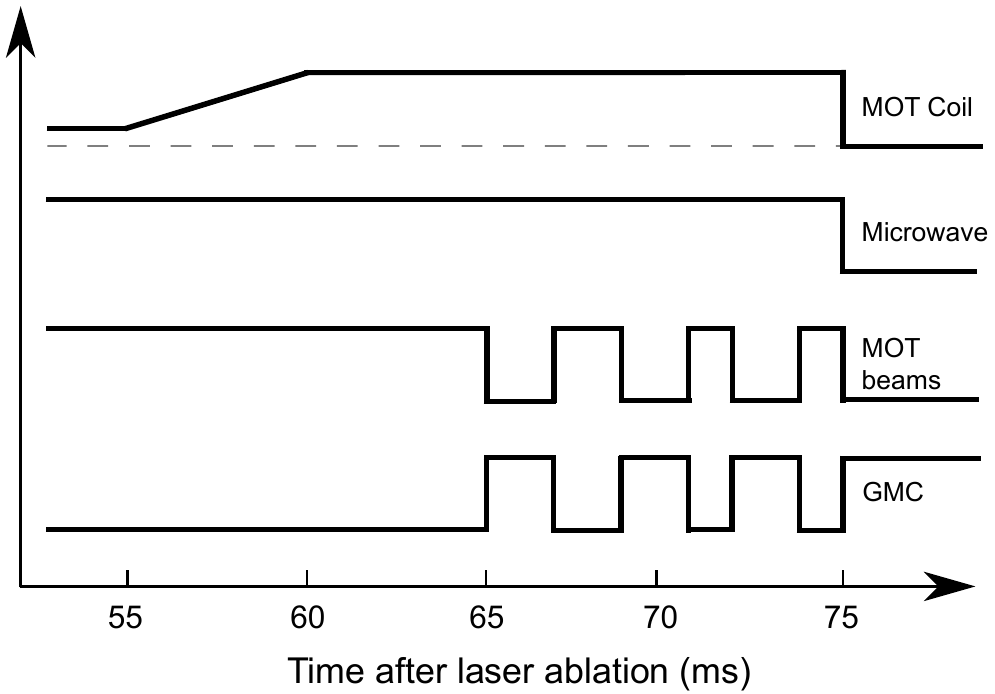}
\caption{\label{fig:timing}
Timing sequence for optimized compression of molecular clouds. 
}
\end{figure}

We first ramp the magnetic field gradient of the DC MOT from 12 G/cm to 47 G/cm in 5 ms, and hold at this value for another 5 ms for initial compression \cite{compressAtom,compressMolecule}. The MOT stabilizes with a diameter of 1.4 mm (2.4 mm) along the axial (radial) direction. 
We then apply a 2 ms GMC pulse to cool the molecules followed by another MOT pulse with a varying duration for compression. The cloud size is measured right after the MOT pulse, as shown in Fig.~\ref{fig:compression}(c). The cloud size shrinks until $\sim$4 ms and is significantly smaller than the equilibrium MOT size. This compression process is dynamical and the cloud expands to eventually reach the original MOT size (Fig.~\ref{fig:compression}(c)).

For our optimized compression sequence, as shown in Fig.~\ref{fig:timing}, we repeat the GMC-MOT cycle three times after the initial 5 ms MOT hold pulse. The GMC pulse is always set at 2 ms, and the MOT pulse follows 2 ms, 1 ms, and 1 ms. The decreasing MOT pulse duration accounts for the shrinking cloud after previous compression cycles. We do not observe considerable molecule loss during the compression except those optically pumped into the unaddressed dark states. 
After the compression, the cloud has a diameter of 0.69 mm (0.93 mm) along the axial (radial) direction. The corresponding volume is approximately 10 times smaller than that before the compression with peak spatial density of $5.4 \times 10^{7} \, \text{cm}^{-3}$ (within a factor of 2). The uncertainty arises mainly from the molecular number estimation.
After the final 1 ms MOT pulse, the molecules are then cooled to 4 $\mu$K with a final 8 ms GMC pulse, creating a molecular sample with peak phase space density of 3.3$\times$10$^{-8}$. 
Both the spatial density and the phase space density are higher than the previous best values in free space~\cite{lambdaehanced}.

\section{Conclusion}

We have investigated a wide range of cooling and trapping strategies for YO molecules based on the unique energy level structure, demonstrating both red-detuned and dual-frequency DC MOTs along with robust, field-insensitive sub-Doppler cooling to 4 $\mu$K with gray molasses cooling. We explored the relation of Raman resonance and GMC and clarified the role of a $\Lambda$-type system in GMC, and demonstrated magnetically induced Sisyphus cooling. In addition, we invented a novel scheme to compress the molecular cloud by combining the MOT spring force with the magnetic-insensitive GMC, which improves the phase space density by one order of magnitude. This scheme may be applicable to a variety of molecules and atoms. Given the insensitivity of GMC to laser detuning (both one photon and two photon), the compressed ultracold molecular cloud can be readily loaded into an optical dipole trap with high efficiency in the presence of cooling, paving the way for investigation of ultracold molecular collisions. It is also feasible to load the molecules into an array of optical tweezers~\cite{moleculetweezer} for quantum information processing~\cite{KaufmanTweezer,DeMille2002QL,Ni2018QL, Preskill2019,Oxford2019}.  

\begin{acknowledgements}
We thank J. Bartolotta, P. He, L. Liu, A. Shankar, A. M, Rey, and M. Holland for stimulating discussions, and A. Collopy, G. Valtolina, P. He, and the J. Doyle group of Harvard for careful reading of the manuscript.\ I. A. F. acknowledges support from the National Research Council postdoctoral fellowship. This work is supported by ARO-MURI, NIST, and NSF Grant No. PHY-1734006.
\end{acknowledgements}

\end{document}